# Neutron diffraction study of magnetism in van der Waals layered $MnBi_{2n}Te_{3n+1}$

Lei Ding[1], Chaowei Hu[2], Erxi Feng[1], Chenyang Jiang[1], Iurii A. Kibalin[3], Arsen Gukasov[3], MiaoFang Chi[4], Ni Ni[2] and Huibo Cao[1,*]

[1] Neutron Scattering Division, Oak Ridge National Laboratory, Oak Ridge, TN 37831, USA
[2] Department of Physics and Astronomy and California NanoSystems Institute, University of California, Los Angeles, CA 90095, USA
[3] Laboratoire Léon Brillouin, CEA, Centre National de la Recherche Scientifique, CE-Saclay, 91191 Gif-sur-Yvette, France
[4] Center for Nanophase Materials Sciences, Oak Ridge National Laboratory, Oak Ridge, TN 37831, USA

[*] E-mail: caoh@ornl.gov

## Abstract

Two-dimensional van der Waals $MnBi_{2n}Te_{3n+1}$ (n = 1, 2, 3, 4) compounds have been recently found to be intrinsic magnetic topological insulators rendering quantum anomalous Hall effect and diverse topological states. Here, we summarize and compare the crystal and magnetic structures of this family, and discuss the effects of chemical composition on their magnetism. We found that a considerable fraction of Bi occupies at the Mn sites in $MnBi_{2n}Te_{3n+1}$ (n = 1, 2, 3, 4) while Mn is no detectable at the non-magnetic atomic sites within the resolution of neutron diffraction experiments. The occupancy of Mn monotonically decreases with the increase of n. The polarized neutron diffraction on the representative $MnBi_4Te_7$ reveals that its magnetization density is exclusively accumulated at the Mn site, in good agreement with the results from the unpolarized neutron diffraction. The defects of Bi at the Mn site naturally explain the continuously reduced saturated magnetic moments from n = 1 to n = 4. The experimentally estimated critical exponents of all the compounds generally suggest a three-dimensional character of magnetism. Our work provides material-specified structural parameters that may be useful for band structure calculations to understand the observed topological surface states and for designing quantum magnetic materials through chemical doping.

## 1. Introduction

The nontrivial topological nature of quantum wavefunctions can give rise to a vast variety of quantum states such as topological insulators (TIs) and quantum Hall effect [1][2][3][4][5][6][7][8][9][10]. Three-dimensional TI is classified by the $Z_2$ topological invariants and featured by a massless Dirac dispersion with the time-reversal invariant due to the spin-orbit interactions [1]. Incorporating spin degree of freedom in TIs in principle leads to the formation of massive Dirac fermions with chiral edge modes, giving way to the occurrence of even novel topological states such as quantum anomalous Hall effect and topological magnetoelectric effects [4][6]. Conventionally, chemical doping of ferromagnetic (FM) metals has been used to generate magnetization in a TI although it normally suffers from inhomogeneity, disorder of dopants and the strict control of the film-fabrication [4][7]. Recently, Mong et al. have theoretically proposed that antiferromagnetic TIs can be realized in systems breaking both time-reversal and translational symmetries but preserving the combination [6]. This, from a symmetry viewpoint, provides a theoretical guide for experimental realization of innate magnetic TIs that are promising to circumvent the obstacles in the conventional magnetic TIs.

A van der Waals layered $MnBi_2Te_4$, characterized by the alternating magnetic Mn-Te layers and non-magnetic Bi-Te layers [11][12][13], was recently proposed to be the first intrinsic antiferromagnetic (AFM) topological insulator [14][15][16]. The atomic arrangement in $MnBi_2Te_4$ naturally compounds the magnetic and topological insulating layers, providing an intrinsic TI material. Later, several independent groups have reported the observation of the Dirac surface states using angle-resolved photoemission spectroscopy (ARPES) [15][16][17][18][19][20][21][22][23][24]. Owing to the nature of van der Waals layers, the structure of $MnBi_2Te_4$ is flexible to incorporate distinct atomic layers. Especially, by increasing the numbers of nonmagnetic $[B_2Te_3]$ building blocks, more members in this family $MnBi_{2n}Te_{3n+1}$ including $MnBi_4Te_7$, $MnBi_6Te_{10}$, $MnBi_8Te_{13}$ and $MnBi_{10}Te_{16}$ were subsequently synthesized and investigated in terms of their magnetic, transport and topological properties [25][26][27][28][29][30][31][32][33][34][35][36]. It has been found that with the increase of n (the number of nonmagnetic $[B_2Te_3]$ building blocks), the magnetic ground state changes from AFM in n = 1, 2, 3 to FM in n = 4 and 5 by suppressing the interlayer AFM exchange interactions [35][36].

Neutron diffraction has shown that $MnBi_2Te_4$ orders with the A-type spin configuration that breaks the product of the time-reversal symmetry and the primitive-lattice translational symmetry at the (001) surface [12][29]. Theoretically, this should cause the formation of a surface Dirac cone with a sizable gap [6]. However, high-resolution ARPES experiments consistently concluded that $MnBi_2Te_4$ exhibits gapless topological surface states below and above the AFM ordering temperature [18][19][20][21][24]. The gapless feature of the topological surface states were also observed on the [MnBi$_2$Te$_4$] terminations in MnBi$_4$Te$_7$ [28] and MnBi$_6$Te$_{10}$ [30][31][32]. This deviation from the theoretical prediction was attributed to either a distinct magnetic configuration at the surface of the materials or disordered Mn distributions [18][20][21][24][30]. Recent static and time-resolved ARPES with full polarization control and magnetic force microscopy studies of the surface layers of MnBi$_2$Te$_4$ have shown that the A-type magnetic structure persists on the surface of the crystals [37][38]. This indicates that the first conjecture about the surface magnetism failed to explain the gapless feature and a comprehensive study of the magnetic defect distributions is crucial.

Apart from studying the origin of the topological surface states, another intriguing aspect is chemical doping which enables the continuous tuning of the magnetism, band topology or carrier density [21][42][43][44]. This, on the other hand, favors the presence of the FM phase and emergence of related topological surface states. A more interesting point will be the realization of the rational tuning of the Fermi level that has been a long-sought goal for the quantum anomalous Hall effect [4]. Indeed, the n–p type carrier transition and the rich magnetic phase diagram including a ferrimagnetic order were reported in Mn(Bi$_x$Sb$_{1-x}$)$_2$Te$_4$ [21][39][41]. A subsequent work of Sb-doping in MnBi$_4$Te$_7$ reveals a more complex evolution of magnetism and interesting transport properties [44]. It has been suggested that the dramatic effect of Sb dopants arises from the Mn/Sb site-mixing defects which in MnSb$_2$Te$_4$ favors a FM interlayer coupling [40][43]. Understanding the site defects in the parent materials will provide a solid base for further efforts on designing magnetic topological materials via chemical doping.

In this work, we summarize and compare the crystal structure and magnetic properties of the family MnBi$_{2n}$Te$_{3n+1}$. Neutron diffraction, energy-dispersive x-ray spectroscopy, and the magnetization density distributions all indicate that only site defects in Mn-layers were observed in this family and should not considerably influence the magnetism within the observed level.

## 2. Experimental methods

Single crystals of MnBi$_{2n}$Te$_{3n+1}$ (n=1, 2, 3, 4) were grown using self-flux method as described in the previous reports [28][29][35]. Single crystal neutron diffraction experiments were performed at the HB-3A DEMAND equipped with a 2D detector at the High Flux Isotope Reactor (HFIR) at ORNL in the temperature range of 4.5-30 K. Neutron wavelengths of 1.008 and 1.553 Å were selected by a bent perfect Si-331 and Si-220 monochromator[45][46]. The rocking curves of peaks were measured by counting the number of scattered neutrons at each rocking angle ω. Rietveld refinements of the crystal and magnetic structures were performed using the Fullprof program [47] against the reduced data. Group-theoretical calculations were done using Bilbao crystallographic server tools [48].

Polarized neutron diffraction was performed with the neutron beam of 1.553 Å polarized by the in-situ pumped $^3$He polarizer. The polarization calibrated by the beam transmission and Heusler crystal is 55% for this experiment. The data were collected at 5 K under a fixed field of 0.83 T provided by a permanent magnet set mounted on the cold head in a closed cycle refrigerator. A total 136 flipping ratios were collected on a MnBi$_4$Te$_7$ crystal. The Cryspy software was used for the flipping ratio analysis and the spin density reconstruction through the maximum entropy method [49][50].

SEM-EDS (Scanning Electron Microscope - Energy Dispersive X-Ray Spectroscopy) analysis was performed on a Hitachi S4800 Cold Field Emission SEM using a voltage of 20kV. ZAF correction method was used for the elemental quantitative analysis.

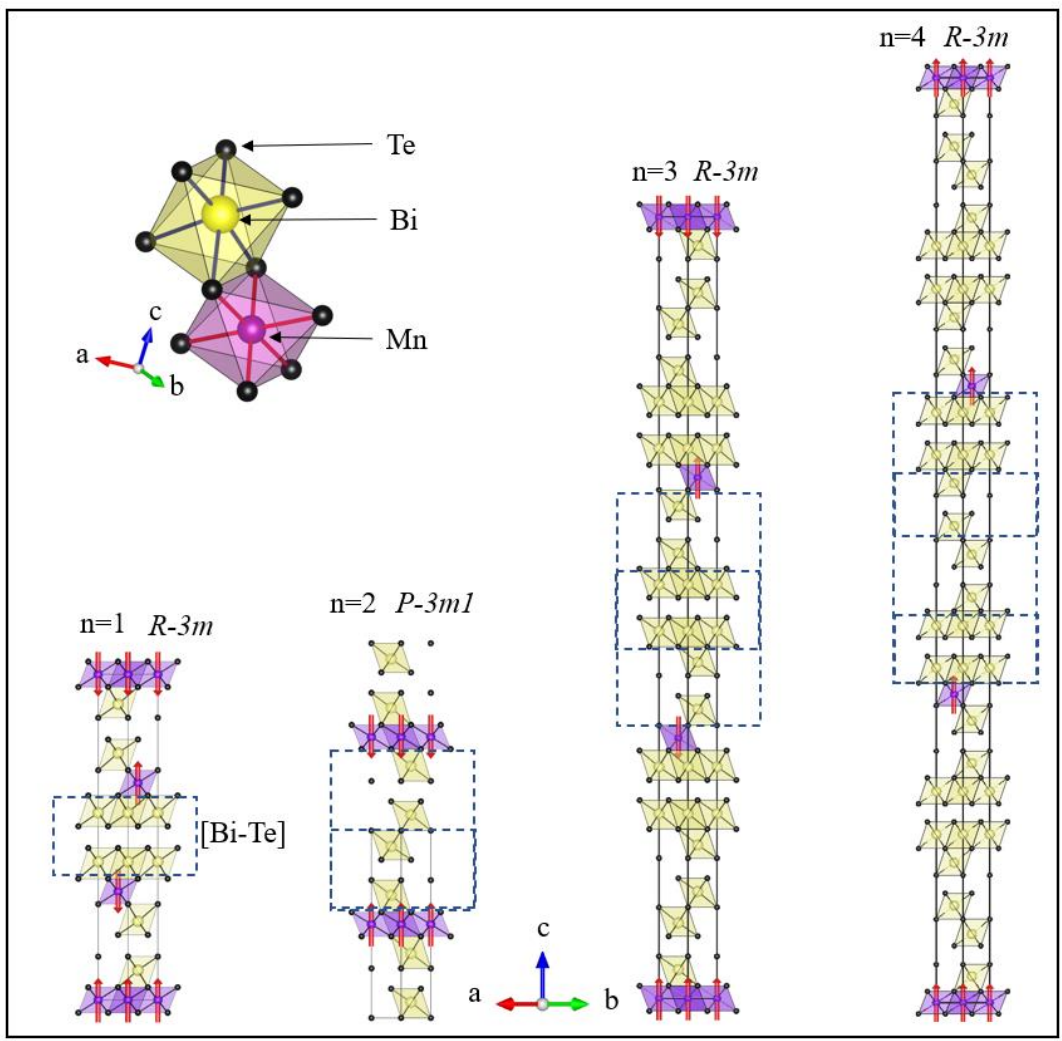

**Figure 1**. Schematic crystal and magnetic structures of MnBi$_{2n}$Te$_{3n+1}$ (n=1, 2, 3, 4). The [Bi-Te] building blocks are marked with rectangles. The unit cell for each case is outlined in solid lines. Top left shows the MnTe$_6$ and BiTe$_6$ octahedra.

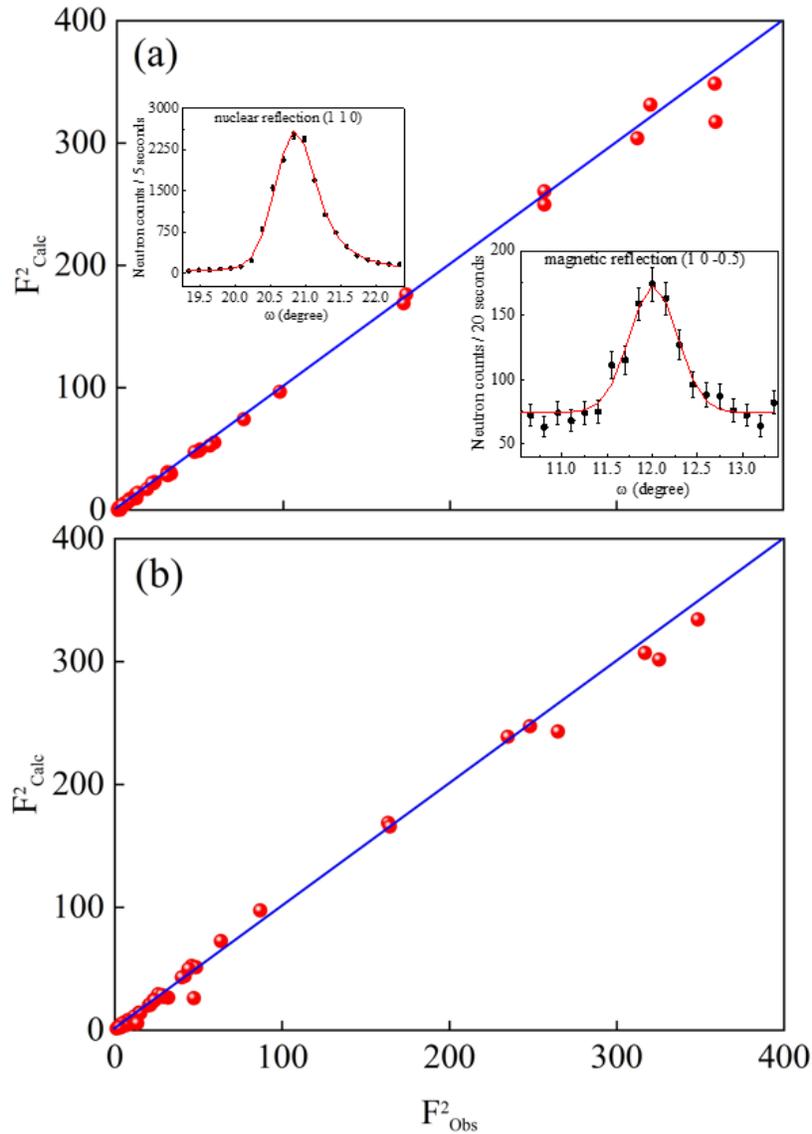

**Figure 2**. (a) Observed and calculated nuclear and magnetic reflections of Mn1-6-10 at 4.5 K. The refinement of the magnetic model gives $R_{F(magnetic)}$ = 7.81%. Inset (left) shows the representative neutron rocking curve scan for the reflection (1 1 0). Inset (right) shows the representative neutron rocking curve scan for the magnetic reflection (1 0 -0.5). (b) Observed and calculated nuclear structure of Mn1-8-13 at 20 K with the agreement factor $R_F$ = 9.11%.

## 3. Results and discussion

The crystal structures of $MnBi_2Te_4$ (Mn1-2-4) and $MnBi_4Te_7$ (Mn1-4-7) have been determined using neutron diffraction technique in our previous work [29]. As shown in figure 1, $MnBi_2Te_4$ and $MnBi_4Te_7$ crystalize with $R$-$3m$ and $P$-$3m$1 space group, respectively, in compliance with the stacking rule proposed for the vdW Mn-Bi-Te series [29]. $MnBi_6Te_{10}$ (Mn1-6-10) crystallizes with the space group $R$-$3m$ with the lattice parameters (20 K) $a = b = 4.361$ Å, $c = 101.300$ Å, $\alpha = \beta = 90°$, $\gamma = 120°$. The nuclear structure of $MnBi_6Te_{10}$ was refined using the neutron diffraction data at 4.5 K. Like $MnBi_2Te_4$ and $MnBi_4Te_7$, the Mn sites are occupied by both Mn and Bi atoms in the ratio 0.66(1): 0.34(1) Mn: Bi in $MnBi_6Te_{10}$. The structural refinement result and the corresponding structural parameters are shown in figure 2(a) and Table 1, respectively. As demonstrated in Figure 1, from Mn1-2-4 to Mn1-6-10, two more non-magnetic [Bi-Te] blocks are inserted making the adjacent Mn layers more distant.

**Table 1.** Refined structural parameters for Mn1-6-10 (n=3) at 4.5 K based on the single crystal neutron diffraction data measured at HB-3A (number of reflections: 105; number of effective reflections: 105 with I > 3σ; $R_F$ = 3.42%, $\chi^2$ = 1.06). The thermal displacement parameter for all the atoms was constrained to be the same in the refinement.

| Atom | Site | x | y | z | $B_{iso}$ (Å$^2$) | Occ. |
|---|---|---|---|---|---|---|
| Mn1/Bi1 | 3a | 0 | 0 | 0 | 0.6(1) | 0.66(1)/0.34(1) |
| Bi2 | 6c | 0 | 0 | 0.23656(3) | 0.6(1) | 1 |
| Bi3 | 6c | 0 | 0 | 0.29633(3) | 0.6(1) | 1 |
| Bi4 | 6c | 0 | 0 | 0.47020(3) | 0.6(1) | 1 |
| Te1 | 6c | 0 | 0 | 0.05387(5) | 0.6(1) | 1 |
| Te2 | 6c | 0 | 0 | 0.11666(4) | 0.6(1) | 1 |
| Te3 | 6c | 0 | 0 | 0.17954(6) | 0.6(1) | 1 |
| Te4 | 6c | 0 | 0 | 0.34939(5) | 0.6(1) | 1 |
| Te5 | 6c | 0 | 0 | 0.41304(4) | 0.6(1) | 1 |

To study the magnetic structure of Mn1-6-10, we have performed neutron diffraction experiments at different temperatures. A set of satellite reflections below 10.3 K emanating from the magnetic phase can be indexed by a vector **k** = (0, 0, 3/2) (figure 2(a)). To solve the magnetic structure of Mn1-6-10, we carried out the magnetic symmetry analysis by considering the **k** vector and its parent space group. There are four corresponding magnetic isotropy subgroups $R_I$-3c (irrep: mT2+), $Cc2/c$ (mT3+), $Cc2/m$ (mT3+) and $P_S$-1 (mT3+). By the symmetry constraint, magnetic space group $R_I$-3c gives only the non-zero $M_Z$ component while $Cc2/m$ allows $M_X$ active and $Cc2/c$ conveys a magnetic configuration ($M_X$, 2$M_X$, $M_Z$). In light of the previous bulk magnetic measurements and the absence of magnetic reflections along the 00L direction [35], the allowed magnetic moment should be along the c axis (i.e. $M_Z$). Therefore, the magnetic space group $R_I$-3c should be an appropriate solution. We refined the corresponding magnetic model against neutron diffraction data at 4.5 K. The refinement result is shown in figure 2(a). The magnetic space group $R_I$-3c delivers a satisfactory result, yielding the A-type magnetic configuration, as shown in figure 1, similar to the spin configuration in MnBi$_2$Te$_4$ and MnBi$_4$Te$_7$ [29]. The refined total magnetic moment at 4.5 K is 3.8(4) $\mu_B$.

The crystal structure of MnBi$_8$Te$_{13}$ (Mn1-8-13) was previously found to crystallize in the $R-3m$ symmetry with the lattice parameters: $a = b = 4.37485(7)$ Å, $c = 132.415(3)$ Å, $\alpha = \beta = 90°$, $\gamma = 120°$ [35]. Here we use the neutron diffraction data at 20 K to refine its nuclear structure. The advantage of high neutron scattering contrast between Mn and Bi atoms, better than x-ray element contrast, enables us to quantitatively determine the atomic occupancy in Mn1-8-13. The structural refinement yields that the non-stoichiometric effect occurs also in MnBi$_8$Te$_{13}$. The Mn sites are occupied by approximately 36(4) % of Bi and 64(4) % of Mn whereas there is minimal Mn residing on the Bi sites. The refinement results and structural parameters are shown and summarized in figure 2(b) and Table 2, respectively. Our previous work has shown that Mn1-8-13 undergoes a ferromagnetic phase transition at 10.5 K [35]. The magnetic structure is described by the magnetic space group $R$-3$m'$ which supports all spins along the $c$ axis. Using the scale factor from the nuclear phase, we calculated the magnetic moment of Mn1-8-13. The intensity of the magnetic reflection (1 0 1) [35] gives the ordered magnetic moment of 4.7(2) $\mu_B$/Mn$^{2+}$ at 4.5 K.

**Table 2**. Refined structural parameters for Mn1-8-13 (n=4) at 20 K based on the single crystal neutron diffraction data measured at HB-3A (number of reflections: 74; number of effective reflections: 64 with I > 4σ; R$_F$ = 2.58%, $\chi^2$ = 2.82). The thermal displacement parameter for all the atoms was constrained to be the same in the refinement.

| Atom | Site | x | y | z | $B_{iso}$ (Å$^2$) | Occ. |
|---|---|---|---|---|---|---|
| Mn1/Bi1 | 3a | 0 | 0 | 0 | 0.16(9) | 0.64(4)/0.36(4) |
| Bi2 | 6c | 0 | 0 | 0.1821(1) | 0.16(9) | 1 |
| Bi3 | 6c | 0 | 0 | 0.22827(9) | 0.16(9) | 1 |
| Bi4 | 6c | 0 | 0 | 0.3620(2) | 0.16(9) | 1 |
| Bi5 | 6c | 0 | 0 | 0.40760(8) | 0.16(9) | 1 |
| Te1 | 6c | 0 | 0 | 0.0412(2) | 0.16(9) | 1 |
| Te2 | 6c | 0 | 0 | 0.0901(4) | 0.16(9) | 1 |
| Te3 | 6c | 0 | 0 | 0.1378(1) | 0.16(9) | 1 |
| Te4 | 6c | 0 | 0 | 0.2719(2) | 0.16(9) | 1 |
| Te5 | 6c | 0 | 0 | 0.3212(2) | 0.16(9) | 1 |
| Te6 | 6c | 0 | 0 | 0.4517(3) | 0.16(9) | 1 |
| Te7 | 3b | 0 | 0 | 0.5 | 0.16(9) | 1 |

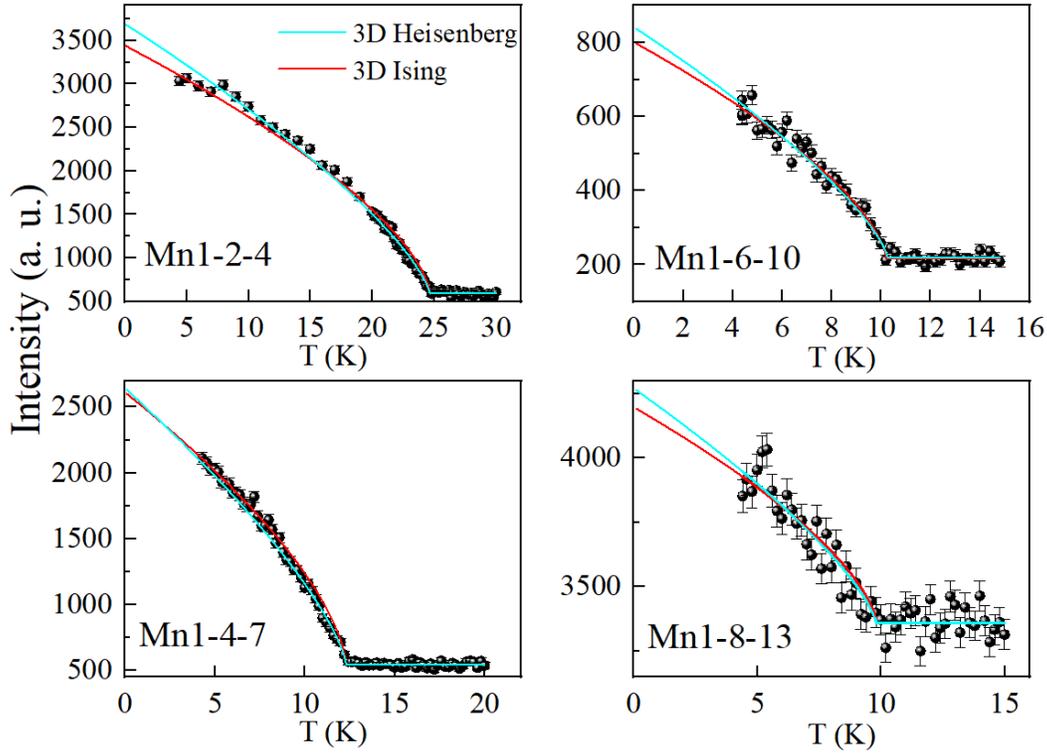

**Figure 3**. Magnetic order parameters upon warming at the magnetic reflections (-1 0 0.5), (0 1 1.5), (0 1 1.5) and (1 0 1) for n=1, n=2 [29], n=3 and n=4 [35], respectively. Solid lines represent the fits using the 3D Heisenberg ($\beta = 0.367$) and 3D Ising ($\beta = 0.326$) models.

The critical behavior of the magnetic phase transitions in $MnBi_{2n}Te_{3n+1}$ can be analyzed using a power-law fit which takes the equation $I = A\left(\frac{T_M - T}{T_M}\right)^{2\beta} + B$, where $T_M$ is the critical temperature for magnetic phase transitions, $A$ is a proportionality constant, $\beta$ is the order parameter critical exponent and $B$ is the background [51][52][53]. For the temperature-dependent magnetic reflections of Mn1-2-4 and Mn1-4-7, attempts using a unique critical exponent to fit the whole data below $T_N$ were unsatisfactory. Therefore, we fitted the data in two temperature ranges using different critical exponents [29]. For Mn1-2-4, the data in the temperature range of 20-30 K can be well fitted using the critical exponent 0.367 that is the value expected for the three dimensional Heisenberg model (3DHM) [51]. In the temperature range of 4-20 K, the data curve, however, follows the critical behavior with the exponent 0.326 for the three dimensional Ising model (3DIM) [52], as shown in figure 3. Hence, the critical behavior in Mn1-2-4 shows a crossover around $0.83T_N$. This behavior was also found in Mn1-4-7 when fitting the intensity of the magnetic reflection of Mn1-4-7 using the power-law. The data above and below 9 K were fitted using the 3DHM and 3DIM, respectively, giving rise to a crossover ~ $0.7T_N$. By contrast, the entire data range of both Mn1-6-10 and Mn1-8-13 can be fitted satisfactorily using only the 3DHM (figure 3), yielding the magnetic ordering temperature 10.3(1) K and 9.8(1) K, respectively. At first glance, a 3D feature of the magnetic model for the whole family compounds is unexpected given the fact that the crystal structure is characterized by the well-separated magnetic layers. However, a recent inelastic neutron scattering work on the powder Mn1-2-4 has disclosed that the magnetic interactions are Ising-like with strong interlayer exchange interactions [54]. This indicates that the interlayer magnetic interactions could be significant even though the inter-magnetic layers are considerably distant. The critical behavior of the 3DIM adopted in Mn1-2-4 and Mn1-4-7 is consistent with the Ising-like nature from the neutron inelastic scattering results [54][55].

The power-law behavior can be extrapolated to the zero-temperature limit for all the cases, allowing us to evaluate the ordered magnetic moment at zero temperature. Following the exponent parameters of the 3DIM and using the magnetic moment from the magnetic structure refinement at 4.5 K, we obtained the ordered magnetic moments 4.9(1) $\mu_B$ and 4.6(1) $\mu_B$ for Mn1-2-4 and Mn1-4-7, respectively. For Mn1-6-10 and Mn1-8-13, the estimated magnetic moments are 4.4(5) $\mu_B$ and 4.9(2) $\mu_B$, respectively. These values of the ordered magnetic moments at zero temperature are close to the expected totally ordered moment of 5 $\mu_B$ for $Mn^{2+}$ ions.

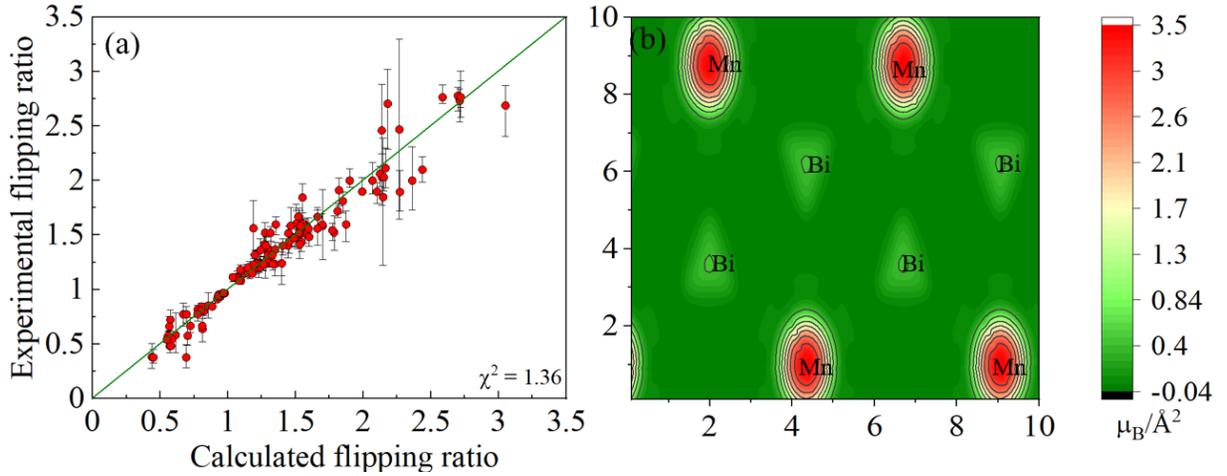

**Figure 4.** (a) The observed and calculated flipping ratios using 136 reflections. (b) Projection of the spin density within the unit cell on the *(ab)*-plane for $MnBi_4Te_7$ under the magnetic field of 0.83 T.

Polarized neutron diffraction is recognized as a unique technique to map out the direct three dimensional distribution of the magnetization in the unit cell [58][59]. As it could further identify the distribution of magnetic impurity, we performed the polarized neutron diffraction experiment on a representative member $MnBi_4Te_7$. The flipping ratio, which stands for the ratio between the spin-up and sip-down intensities of the Bragg reflections, can be expressed as follows:

$$R = \frac{I_+}{I_-} = \frac{F_N^2 + 2p\, sin\alpha^2 F_N F_M + sin\alpha^2 F_M^2}{F_N^2 - 2p\, e\, sin\alpha^2 F_N F_M + sin\alpha^2 F_M^2} \qquad (1)$$

where $F_N$ and $F_M$ designate the nuclear and magnetic structure factors, respectively; α is the angle between the magnetic field direction and the scattering vector, e is the flipping efficiency, $p = 0.55$ is the polarization of the beam. This expression is simplified for centrosymmetric structure, which allows us to calculate the Fourier components of the magnetization density. The nuclear structure determined by the previous unpolarized neutron diffraction was used in the calculations [29]. As shown in Figure 4 (a), the refinement using the nuclear structure yields a magnetic moment of 4.2(2) $\mu_B$, in good agreement with that determined from the unpolarized neutron diffraction. Since the unpolarized neutron diffraction revealed a small amount of Mn, 1(1)%, at the Bi sites [29], we also evaluate the site defects using the polarized neutron data. The reconstructed magnetization density map from the Cryspy software shows that the magnetization density is exclusively accumulated around the Mn site (1$b$ site) which corresponds to a magnetic moment of 4.2(2) $\mu_B$. We did not observe visible magnetization density around the Bi sites, indicating that there is no Mn content distributed at the Bi sites within the resolution of our experiment.

Having known the crystal structures of $MnBi_{2n}Te_{3n+1}$, we can compare their structural parameters to establish the relationship between structure and magnetic properties. Figure 5 (a) shows the principle bond lengths for $MnBi_{2n}Te_{3n+1}$ (n = 1, 2, 3, 4). With the increase of n (the number of building blocks), the averaged Bi-Te bond length in the octahedron monotonically increases. This tendency is generally similar to the change of lattice constant $a$. We examined the change of the Te-Te bond lengths in the adjacent layers in which the magnetic layers reside. As shown in figure 5 (b), the Te-Te distance first increases from n = 1 to n = 3 then drops to a lower value for n = 4. This implies that, in stabilizing the $MnBi_8Te_{13}$ structure, there may exist a small strictive effect in this layer to properly incorporate the four building blocks between the magnetic layers [56]. The anomaly may be associated with the change of topological properties from n = 3 to n = 4. A recent work about the stacking-dependent topological property on the nonmagnetic isostructural $MBi_2Te_4$ (M=Ge, Sn, Pb) has emphasized the important role of the interlayer Te-$p_z$ orbital coupling and the "Te quasicovalent" bonding in manifesting different topological properties [57]. The observation of the sudden drop of interlayer Te bond distance from n = 3 to n = 4 may indicate the change of the interlayer Te-$p_z$ orbital coupling. This seems to be in accordance with the variation of the topological states as suggested in $MnBi_6Te_{10}$ [31] and $MnBi_8Te_{13}$ [35]. Furthermore, this is likely associated with the switch of interlayer interactions (from AFM to FM), indicative of the coupling between topology and magnetism.

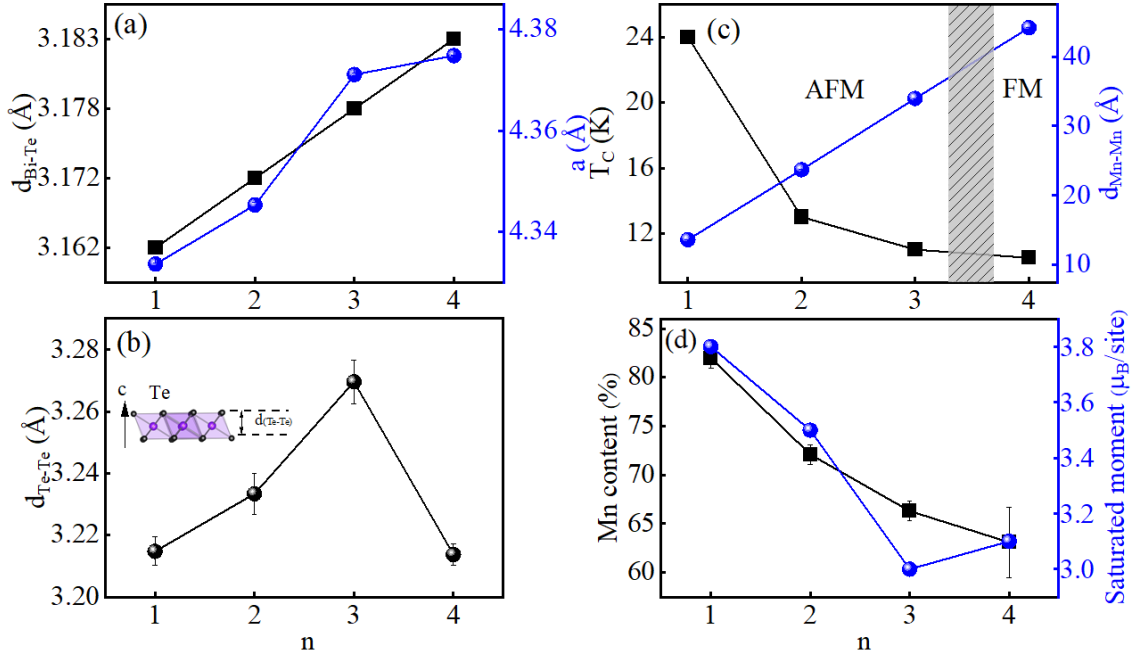

**Figure 5.** (a) The average Bi-Te bond length in the BiTe$_6$ octahedron and lattice constant $a$ as a function of n. (b) The distance between the nearest Te layers close to the Mn site as a function of n. Inset shows the distance between the neighboring Te cations. (c) The ordering temperatures and the Mn-Mn distance between adjacent layers in MnBi$_{2n}$Te$_{3n+1}$ (n=1, 2, 3, 4). The hatched zone represents the unknown magnetic ground states. (d) The occupancies of Mn and their saturated magnetic moments at 2 K with the field along the $c$ axis for MnBi$_{2n}$Te$_{3n+1}$ (n=1, 2, 3, 4).

As the magnetism in MnBi$_{2n}$Te$_{3n+1}$ originates from the exchange interactions between the Mn atoms, we appreciate the Mn-Mn distance between the adjacent layers. As expected, the Mn-Mn distance is linearly proportional to the value of n (figure 5(c)). The tendency of ordering temperatures reasonably goes along with this change but not linearly. MnBi$_8$Te$_{13}$ comes out as a special case as it orders ferromagnetically below 10.5 K. Such a ferromagnetic transition reflects the significant suppression of the antiferromagnetic interaction between Mn atoms in the adjacent layers. For all the cases, the saturated magnetic moments per Mn from the magnetization measurements are significantly smaller than the theoretical values for Mn$^{2+}$ ions. Furthermore, from Mn1-2-4 to Mn1-6-10, the saturated magnetic moment monotonically decreases with the increasing number of the building blocks. Such a trend can be understood by the corresponding increase of the Mn site defects. As shown in figure 5(d), the occupancy of Mn monotonically decreases with the increase of n. This naturally explains the reduced saturated magnetic moments. To have a cross check of chemical composition, we have also performed energy-dispersive x-ray spectroscopy (EDS) experiment on n=1, 2, 3 with several selected zones exposed (see figure S1-S3 in the supporting information). The averaged chemical compositions for these members are reasonably close to the neutron results (Table 3). Note, our EDS indicates Mn1-2-4 has the formula Mn$_{0.86}$Bi$_{2.06}$Te$_4$, consistent with the reported Mn$_{0.85(3)}$Bi$_{2.10(3)}$Te$_4$ [22]. But the fraction of Bi varies by over 20% depending on selected regions.

The large variation of Bi fraction also occurs for Mn1-4-7 and Mn1-6-10. While the consistent Mn/Te ratios from EDS agree with our neutron diffraction results.

**Table 3**. Comparison of the chemical compositions determined by the neutron diffraction and EDS method for compounds $MnBi_{2n}Te_{3n+1}$ (n=1, 2, 3, 4).

| Compounds | Mn (neutron) | Mn (EDS) | Bi (neutron) | Bi (EDS) | Te (neutron) | Te (EDS) |
|---|---|---|---|---|---|---|
| n=1 | 0.82(1) | 0.85-0.87 | 2.18(1) | 2.00-2.56 | 4 | 4 |
| n=2 | 0.72(1) | 0.71-0.76 | 4.28(1) | 3.44-4.03 | 7 | 7 |
| n=3 | 0.66(1) | 0.70-0.77 | 6.34(1) | 5.68-6.33 | 10 | 10 |
| n=4 | 0.64(4) | - | 8.36(4) | - | 13 | - |

Previous results based on x-ray diffraction and electron microscopy on the nominal $MnBi_2Te_4$ have revealed the composition $Mn_{0.85(3)}Bi_{2.10(3)}Te_4$ with the presence of anti-site disorder between Mn and Bi sites and Mn vacancies in the nominal $MnBi_2Te_4$ [22]. Similar defects have been also reported in the nominal $MnBi_4Te_7$ by transmission electron microscopy [25]. In contrast to these techniques, neutron diffraction technique has its advantages since Mn and Bi atoms have opposite sign of the neutron scattering length. Our finding of the presence of site defects in this family of van der Waals compounds indicates that the site defects are inherent and should be a common mark to this kind of materials. In fact, when considerable Mn defects occur at the Bi sites, the magnetic properties will be principally modified. In some cases, spontaneous magnetization should take place if the anti-site defects are prevailing. A relevant example is the isostructural $MnSb_2Te_4$ [40][43]. Even though it is similar in structure with $MnBi_2Te_4$, $MnSb_2Te_4$ was found to show spontaneous magnetization. Neutron diffraction reveals that $MnSb_2Te_4$ is ferrimagnetic at low temperatures. Combining theoretical calculations, it was found that the presence of site mixing between Sb and Mn sites plays an important role for the interlayer Mn-Mn ferromagnetic interactions as the site mixing alters the interlayer exchange coupling from antiferromagnetic to ferromagnetic [40]. With regards to the great difference between Mn and Bi in electronegativity, the occurrence of anti-site defects in $MnBi_{2n}Te_{3n+1}$ are basically unfavorable [61]. Due to the strong electronegativity of Bi, the presence of Bi defects at the Mn sites in such a large unit cell is likely to stabilize the structural arrangement, for instance, via the *p*-orbital chemical bonding mediated by the *p* orbitals from Te atoms [60]. It looks plausible when considering the monotonical increase of the Bi defects at the Mn sites from n=1 to n=4. The 3-dimensional magnetization density distribution from the polarized neutron diffraction further excludes the presence of Mn at Bi and Te sites.

Stacking faults that manifest as diffuse rods are normally prevailing in layered materials where translations of the layers parallel to their own planes or to irregular sequence of layers of different kinds are present [62][63][64][65]. In view of the rather weak bonding (van der Waals) between adjacent triangular layers in MnBi$_{2n}$Te$_{3n+1}$, stacking faults are expected to occur. However, in the course of studying the crystal structure of the MnBi$_{2n}$Te$_{3n+1}$ family using neutron diffraction, we found that the stacking faults are minimal in all the studied cases. In contrast to the randomly distributed magnetic dopants in the conventional doped magnetic topological insulators systems [4][7], the magnetism in the family of MnBi$_{2n}$Te$_{3n+1}$ is intrinsic. These facts altogether reflect that MnBi$_{2n}$Te$_{3n+1}$ is a set of clean systems spanning antiferromagnetic and ferromagnetic order to study the interplay between topological states and magnetism.

## 4. Conclusions

The crystal structures and magnetic properties of MnBi$_{2n}$Te$_{3n+1}$ (n = 1, 2, 3, 4) have been systematically studied. The presence of Bi defects at the Mn site rather than the anti-site disorder was found in all the members, reflecting that the formation of Bi defects is inherent and probably independent on the synthesis condition. This is further corroborated by our polarized neutron diffraction results. The elucidation of the presence of site defects only at the Mn sites in the family of compounds may provide insights in understanding the nature of the topological surface states emanating from the magnetism.

**Acknowledgments**

The research at Oak Ridge National Laboratory (ORNL) was supported by the U.S. Department of Energy (DOE), Office of Science, Office of Basic Energy Sciences, Early Career Research Program Award KC0402010, under Contract DE-AC05-00OR22725. This research used resources at the High Flux Isotope Reactor, a DOE Office of Science User Facility operated by ORNL. Work at UCLA was supported by the U.S. Department of Energy (DOE), Office of Science, Office of Basic Energy Sciences under Award Number DE-SC0011978. SEM-EDS was conducted at the Center for Nanophase Materials Sciences, which is a DOE Office of Science User Facility. This manuscript has been authored by UT-Battelle, LLC under Contract No. DE-AC05-00OR22725 with the U.S. Department of Energy. The United States Government retains and the publisher, by accepting the article for publication, acknowledges that the United States Government retains a non-exclusive, paidup, irrevocable, world-wide license to publish or reproduce the published form of this manuscript, or allow others to do so, for United States Government purposes. The Department of Energy will provide public access to these results of federally sponsored research in accordance with the DOE Public Access Plan(http://energy.gov/downloads/doepublic-access-plan).

# Neutron diffraction study of magnetism in van der Waals layered MnBi$_{2n}$Te$_{3n+1}$

Lei Ding[1], Chaowei Hu[2], Erxi Feng[1], Chenyang Jiang[1], Iurii A. Kibalin[3], Arsen Gukasov[3], MiaoFang Chi[4], Ni Ni[2] and Huibo Cao[1,*]

[1] Neutron Scattering Division, Oak Ridge National Laboratory, Oak Ridge, TN 37831, USA
[2] Department of Physics and Astronomy and California NanoSystems Institute, University of California, Los Angeles, CA 90095, USA
[3] Laboratoire Léon Brillouin, CEA, Centre National de la Recherche Scientifique, CE-Saclay, 91191 Gif-sur-Yvette, France
[4] Center for Nanophase Materials Sciences, Oak Ridge National Laboratory, Oak Ridge, TN 37831, USA

[*] E-mail: caoh@ornl.gov

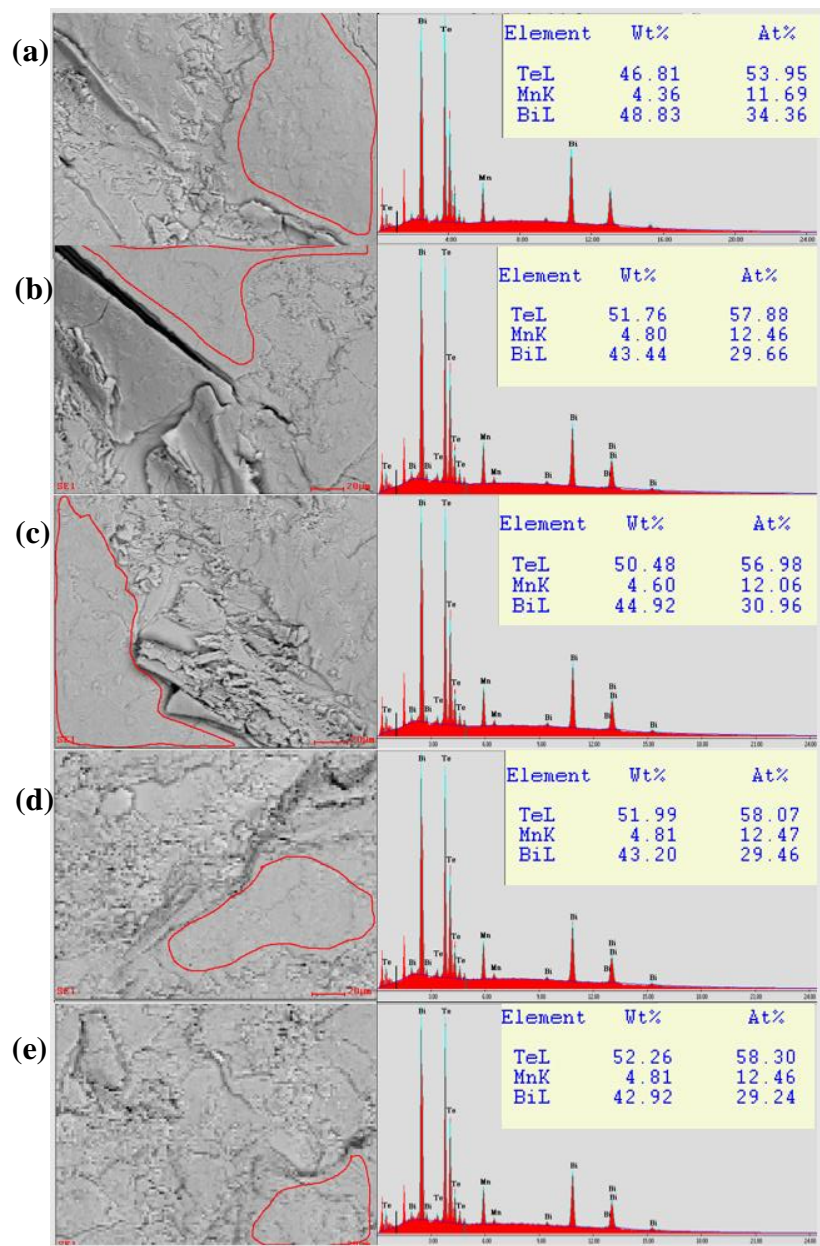

**Figure. S1.** SEM images of the selected regions as well as the corresponding EDS spectrum on MnBi$_2$Te$_4$. The solid line zone denotes the area for the EDS measurement.

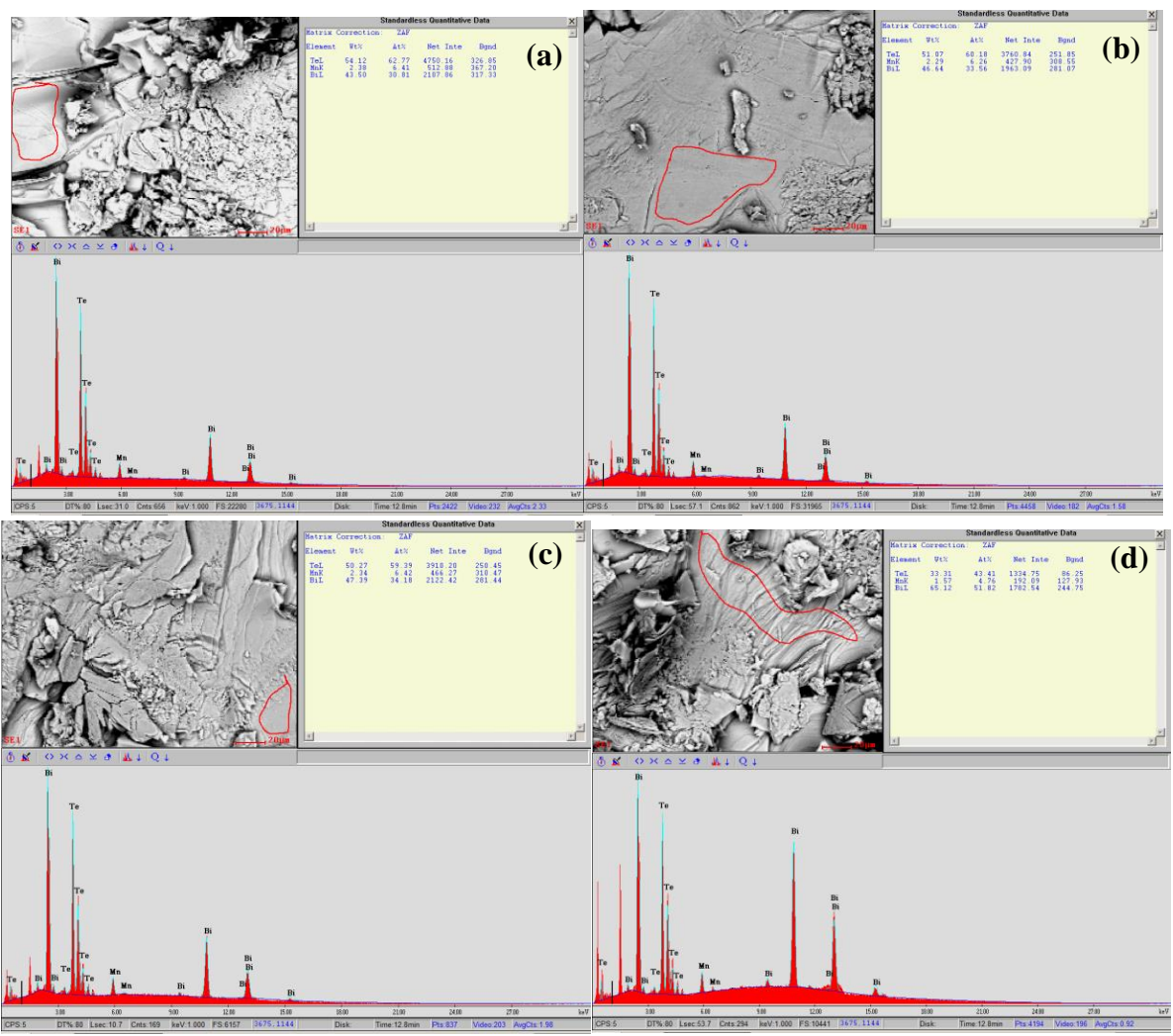

**Figure. S2.** SEM images of the selected regions as well as the corresponding EDS spectrum on MnBi$_4$Te$_7$. The solid line zone denotes the area for the EDS measurement. The SEM image as shown in (d) is not considered.

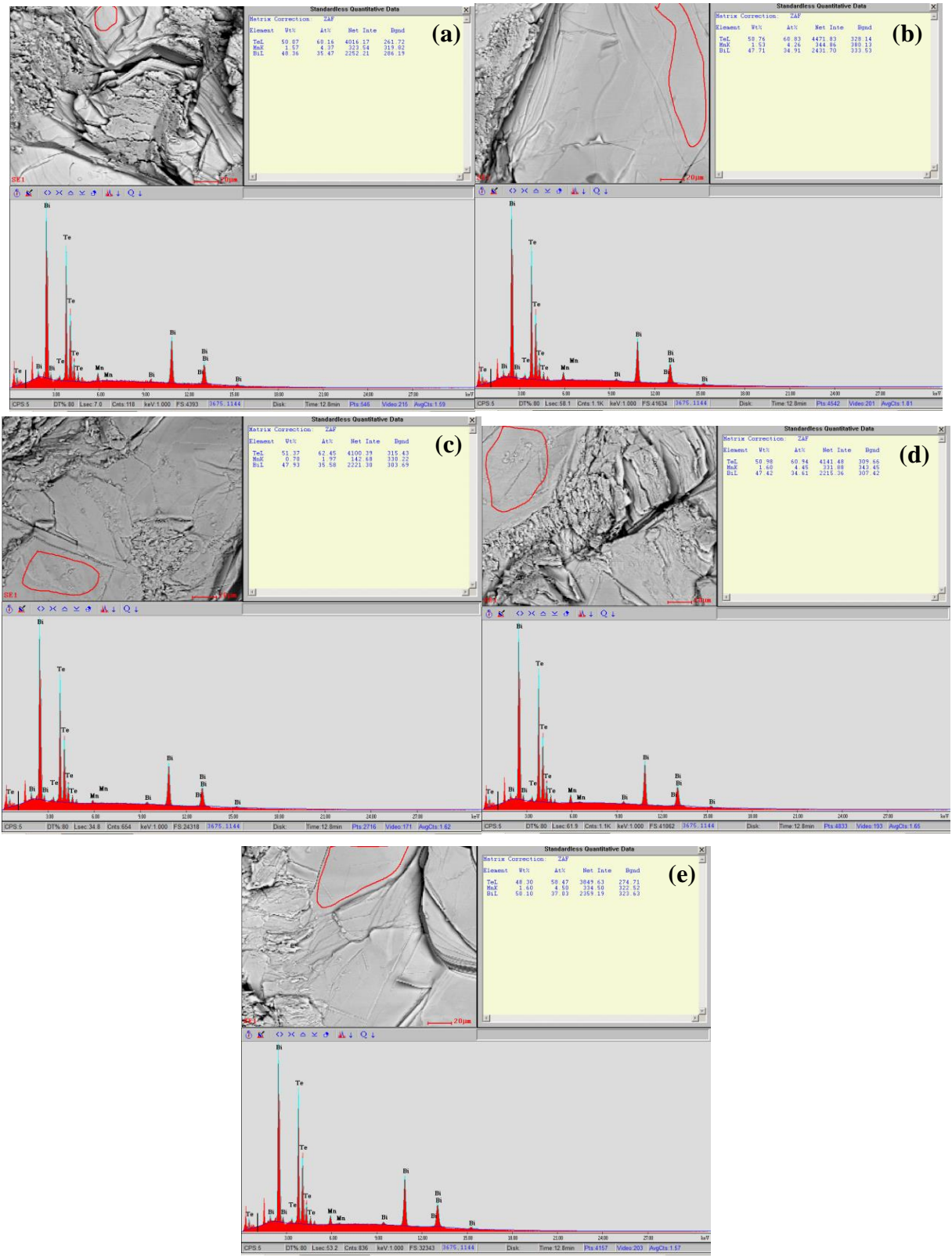

**Figure. S3.** SEM images of the selected regions as well as the corresponding EDS spectrum on MnBi$_6$Te$_{10}$. The solid line zone denotes the area for the EDS measurement. The SEM image as shown in (c) is not considered.